\documentclass[11pt]{article}

\usepackage{color}               
\usepackage{a4wide}
\usepackage{graphicx}
\usepackage{amssymb, amsthm}
\usepackage{epstopdf}
\usepackage{amsmath}
\usepackage{hyperref}

\newtheorem{cor}{Corollary}
\newtheorem{theorem}{Theorem}
\newtheorem{lemma}{Lemma}
\newtheorem{claim}{Claim}

\newcommand{\ds}{\displaystyle}
\newcommand{\OPT}{\textrm{OPT}}

\title{Minimum Entropy Orientations}

\author{
Jean Cardinal\thanks{Universit\'e Libre de Bruxelles, D\'epartement d'Informatique, c.p.~212, B-1050 Brussels, Belgium, jcardin@ulb.ac.be.}
 \and 
Samuel Fiorini\thanks{Universit\'e Libre de Bruxelles, D\'epartement de Math\'ematique, c.p. 216,  B-1050 Brussels, Belgium,  sfiorini@ulb.ac.be.}
\and 
Gwena\"el Joret\thanks{Universit\'e Libre de Bruxelles, D\'epartement d'Informatique, c.p.~212,  B-1050 Brussels, Belgium, gjoret@ulb.ac.be. G. Joret is a Research Fellow of the Fonds 
National de la Recherche Scientifique (F.R.S.--FNRS).}
}

\date{}

\begin{document}
\sloppy

\maketitle

\begin{abstract}
We study graph orientations that minimize the entropy of the in-degree sequence. We prove that the minimum entropy orientation problem is NP-hard even if the graph is planar, and that there exists a simple linear-time algorithm that returns an approximate solution with an additive error guarantee of 1 bit. 
\end{abstract}

{\noindent {\bf Keywords:}
Approximation algorithm; Entropy; Graph orientation; Vertex cover 
}

\section{Introduction}

All graphs considered here are finite, undirected and loopless, but multiple edges are allowed. Let $G=(V,E)$ be a graph with $n$ vertices and $m$ edges, and consider any directed graph $\vec{G}$ obtained by orienting the edges of $G$. The {\em in-degree distribution} $p \in \mathbb{Q}_+^V$ of this orientation is defined by $p_v := \rho_{\vec{G}}(v) / m$, where $\rho_{\vec{G}}(v)$ denotes the in-degree of $v$ in $\vec{G}$.

In this paper, we consider the problem of finding an orientation whose corresponding in-degree distribution is as unbalanced as possible. As a balance measure, we use the {\em entropy}
$$
H(p):=\sum_{v \in V} - p_v \log p_v,
$$
where $\log$ denotes the base $2$ logarithm, and $-0 \log 0 := 0$. The {\em minimum entropy orientation} problem (MINEO) is the problem of finding an orientation of $G$ with an in-degree distribution $p$ minimizing $H(p)$.

The study of MINEO is motivated by that of the {\em minimum entropy set cover} problem (MINESC), introduced by Halperin and Karp~\cite{HK05}. In the latter problem, we are given a ground set $U$ and a collection $\mathcal{S} = \{S_1,\ldots,S_q\}$ of subsets of $U$ whose union is $U$, and we have to assign each element of $U$ to a subset $S_i$ containing it. This assignment partitions $U$ into classes $U_{1}, U_{2}, \dots, U_{q}$ of elements assigned to the same subset, and the objective is to minimize the entropy of the probability distribution $\frac{|U_{i}|}{|U|}$ 
defined by this partition. Hence, MINEO is the special case of MINESC where each element of the ground set can be covered by exactly two sets from $\mathcal{S}$.
(To see this, take $U := E$ and let $\mathcal{S} := \{ S_{v} \mid v \in V\}$ with $S_{v} := \{e \in E \mid e 
\textrm{ is incident to } v\}$.)

The main motivation of Halperin and Karp for introducing MINESC was to solve a haplotyping problem, of importance in computational biology. This problem involves covering a set $U$ of {\em partial haplotypes} of length $d$, defined as words in the set $\{0,1,*\}^d$, by {\em complete haplotypes}, defined as words in $\{0,1\}^d$. Each subset of $\mathcal S$ corresponds to a complete haplotype $h \in \{0,1\}^d$ and contains all partial haplotypes that are {\em compatible} with $h$, that is, partial haplotypes whose symbols match in every non-`$*$' positions with $h$. The `$*$' positions in the partial haplotypes are interpreted as measurements error. It is shown that, under some probabilistic assumptions, minimizing the entropy of the covering amounts to maximizing its likelihood~\cite{HK05}. 

An application of MINEO is the special case of partial haplotyping
in which each partial haplotype has at most one `$*$' in it. Partial haplotypes are then edges of a $d$-dimensional hypercube, and the subsets $S_i$ correspond to vertices of this hypercube. In other words, this special case of the partial haplotyping problem is a minimum entropy orientation problem in a partial hypercube.

\begin{figure}
\begin{center}
\includegraphics[scale=2]{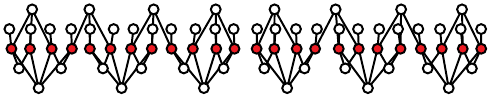}
\end{center}
\caption{\label{fig:tight}Construction of tight examples for the greedy  algorithm. Given a value $t$ (in our illustration, $t = 4$), let $S$ be a set of $t!$ independent vertices (the red vertices). Then for each $i$ in $\{1,2,\ldots ,t\}$, construct $t! / i$ independent vertices of degree $i$ with neighbors in $S$, such that their neighborhoods partition $S$ into subsets of size $i$. The optimal solution is obtained by orienting each edge toward its endpoint in $S$. The greedy algorithm may orient each edge in the opposite direction. This example is equivalent to the tight example previously given for the minimum entropy set cover problem~\cite{CFJ08}.}
\end{figure}

The well-known greedy algorithm for the set cover problem is applicable to MINEO. It involves iterating the following steps: choose a maximum degree vertex $v$ in $G$, orient all edges incident to $v$ toward $v$, and remove $v$ from $G$. The performance of the greedy algorithm for MINESC has been studied thoroughly. Halperin and Karp~\cite{HK05} first showed that the greedy algorithm approximates MINESC to within some additive constant.  Then the current authors improved their analysis and showed that the greedy algorithm returns a solution whose entropy is at most the optimum plus $\log e$ bits, with $\log e\approx 1.4427$ bits~\cite{CFJ08}. Moreover, they proved that it is NP-hard to approximate MINESC to within an additive error of $\log e - \epsilon$, for every constant $\epsilon > 0$. Since MINEO is a special case of MINESC, the first result implies that the greedy algorithm also approximates MINEO to within an additive error of $\log e$ bits. We note that there exist instances of MINEO where the latter bound is (asymptotically) attained; see for example those described in Figure~\ref{fig:tight}.

In this paper, we first prove that MINEO is NP-hard, even if the input
graph is planar (Section~\ref{sec-hardness}). The reduction is from a restricted version of the $1$-in-$3$ Satisfiability problem. Then, we show in Section~\ref{sec-approx} that there exists a simple linear-time approximation algorithm for MINEO with an improved approximation guarantee of $1$ bit. 

To conclude the introduction, we mention that MINEO is also related to
the {\em minimum sum vertex cover} problem (MINSVC), introduced by Feige, Lov{\'a}sz, and Tetali~\cite{FLT04}. In that problem, we are given a graph $G$, and have to find an ordering $v_{1}, v_{2}, \dots, v_{n}$ of its vertices such that the average cover time of an edge is minimized, that is, minimizing
$$
\sum_{i=1}^{n} i \cdot |f(v_{i})|,
$$
where $f(v_{i})$ denotes the set of edges that are incident to $v_{i}$ but to no vertex $v_{j}$ with $j < i$. This problem can also be seen as a graph orientation problem in which the most unbalanced orientation is sought, although with a different balance measure than that of MINEO. Feige {\em et al.}~\cite{FLT04} proved that MINSVC is APX-hard, and gave a $2$-approximation algorithm, based on randomized rounding of a natural linear programming relaxation. Using a different rounding technique, Berenholz, Feige, and Peleg~\cite{BFP06} recently derived an improved approximation factor of $1.99995$.

Although MINEO and MINSVC share some common properties, the main difference between these two problems is perhaps how they behave with respect to instances that are the union of smaller ones: As noted by Berenholz {\em et al.}~\cite{BFP06}, MINSVC is not ``linear'', in the sense that an optimal solution to the union of two disjoint graphs $G_{1}$ and $G_{2}$ is not necessarily a combination of an optimal solution to $G_{1}$ and $G_{2}$, respectively. (In particular, the APX-hardness proof of Feige {\em et al.}~\cite{FLT04} relies on this non-linearity.) On the other hand, MINEO {\em is} linear, as can be easily checked.

\section{Hardness}
\label{sec-hardness}

Let $a=(a_1, a_2, \ldots , a_n)$ and $b=(b_1,b_2,\ldots ,b_n)$ be two sequences of non-negative integers sorted in non-increasing order, and such that $\sum_{i=1}^n a_i=\sum_{i=1}^n b_i =: m$. We say that sequence $a$ {\em dominates} sequence $b$ if 
\begin{equation}
\label{eq-dominance}
\sum_{j=1}^i a_j \geq \sum_{j=1}^i b_j
\end{equation}
for every $i\in \{1, \dots, n\}$, and moreover~(\ref{eq-dominance}) holds with strict inequality for at least one such $i$. We emphasize that $a \neq b$ whenever $a$ dominates $b$. The following lemma is a standard consequence of the strict concavity of the function $x \mapsto -x\log x$; see e.g.~\cite{MEC-ISAAC05, FHM-soda, HLP88} for different proofs. 

\begin{lemma}
\label{lem:jungle}
If $a$ dominates $b$, then $H(a/m) < H(b/m)$. 
\end{lemma}

\begin{theorem}
\label{th-NPhard}
Finding a minimum entropy orientation of a planar graph is NP-hard.
\end{theorem}
\begin{proof}
In the $1$-in-$3$ Satisfiability problem, we are given a 3-SAT formula in input, and we have to decide whether there exists a truth assignment of the variables such that each clause is satisfied by {\em exactly} one of its three literals. Moore and Robson~\cite{MR01} proved that this problem is NP-complete, even if every variable appears in exactly three clauses, there is no negation in the formula, 
and the bipartite graph obtained by linking a variable and a clause if and only if the variable appears in the clause, is planar.

We will reduce the latter restriction of the $1$-in-$3$ Satisfiability problem to MINEO. It will be convenient for the proof to restate Moore and Robson's result in the context of the Exact Cover problem. The latter asks, given a set system $(U, \mathcal{S})$, to decide if $U$ can be covered using pairwise disjoint sets from $\mathcal{S}$. As noted by Li and Toulouse~\cite{LT06}, the NP-completeness of the version of $1$-in-$3$ Satisfiability described above directly implies that Exact Cover is NP-complete even when every set in $\mathcal{S}$ has cardinality exactly 3, each element in $U$ is included in exactly three sets of $\mathcal{S}$, and the ``elements versus sets'' incidence graph is planar. (To see this, consider the set system $(U, \mathcal{S})$ where $U$ and $\mathcal{S}$ are associated with the clauses and the variables of the $1$-in-$3$ Satisfiability instance, respectively.) Let $(U, \mathcal{S})$ be any such set system, and denote by $u_{1}, \dots, u_{q}$ and $S_{1}, \dots, S_{q}$ the elements of $U$ and the sets in $\mathcal{S}$, respectively.
We may assume without loss of generality that $q$ is a multiple of 3, since otherwise
 $(U, \mathcal{S})$ has no exact cover.

We construct a graph $G=(V,E)$ as follows: First, create a vertex $s_{i}$ per set $S_{i}$. Then, for each element $u_{j}$, add a copy of the gadget depicted in Figure~\ref{fig-point-gadget}, and link $u_{j,k}$ ($1\leq k \leq 3$) to $S_{j_{k}}$, where $j_{1}, j_{2},j_{3}$ are the indices of the three sets containing $u_{j}$. 
\begin{figure}
\centering
\includegraphics[width=4.5cm]{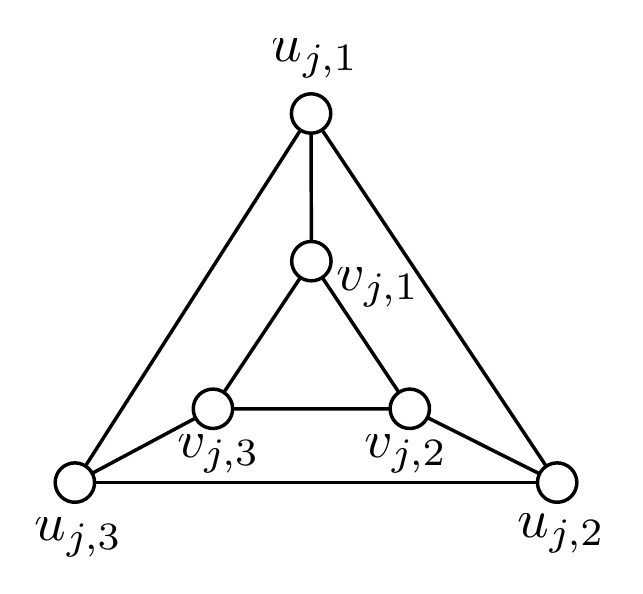}
\caption{\label{fig-point-gadget}Gadget for element $u_{j}$.}
\end{figure}
The fact that $G$ is planar directly follows from the planarity of the bipartite graph underlying the set system $(U, \mathcal{S})$. 

Let $\vec{G}$ be any orientation of $G$ with minimum entropy, and denote by $A$ its arc set. For a subset $X \subseteq V$ of vertices,
we use $\delta(X)$ for the number of arcs going from $X$ to $V - X$ in $\vec{G}$. We define  the {\em in-degree sequence} of $X$, denoted $\textrm{in-seq}(X)$, as the sequence of in-degrees of the vertices in $X$, sorted in non-increasing order. Let also $X_{j}:=\{u_{j,1},u_{j,2}, u_{j,3}, v_{j,1}, v_{j,2}, v_{j,3}\}$, for every $j \in \{1, \dots, q\}$.

\begin{claim}
\label{claim-gadget}
The in-degree sequence of the set $X_{j}$ in $\vec G$ is given by the following table:
$$
\begin{array}{c|c}
\delta(X_{j}) & \textrm{in-seq}(X_{j}) \\
\hline
0 & (4,3,3,1,1,0) \\
1 & (4,3,3,1,0,0) \\
2 & (4,3,2,1,0,0) \\
3 & (3,3,2,1,0,0) \\
\end{array}
$$
\end{claim}
\begin{proof}
Denote by $(c_{1}, c_{2}, \dots, c_{6})$ the in-degree sequence of the set $X_{j}$ in $\vec G$. Arguing by contradiction, we suppose that this sequence is different from the one given in the table.  Considering the structure of the element-gadget, and in particular its two disjoint triangles, we infer the following bounds:
\begin{align*}
c_{1} &\leq \left\{ 
\begin{array}{ll}
4 & \quad \textrm{if } \delta(X_{j}) < 3; \\
3 & \quad \textrm{otherwise};
\end{array}
\right.  \\
c_{2} &\leq 3; \quad
c_{3} \leq 3; \quad
c_{4} \geq 1;\\
c_{5} &\geq 1 \quad \quad \textrm{ if } \delta(X_{j})=0. \\
\end{align*}
It follows from the inequalities above and $\sum_{\ell=1}^6 c_\ell = 12 - \delta(X_j)$ that $\textrm{in-seq}(X_{j})$ is dominated (in the sense of Lemma~\ref{lem:jungle}) by the corresponding sequence in the table. Now, it is always possible to re-orient the arcs of $\vec G$ with both endpoints in $X_{j}$ in such a way that  $\textrm{in-seq}(X_{j})$ realizes the latter sequence, as illustrated in Figure~\ref{fig-gadget-orientation}. 
\begin{figure}
\centering
\includegraphics[width=5cm]{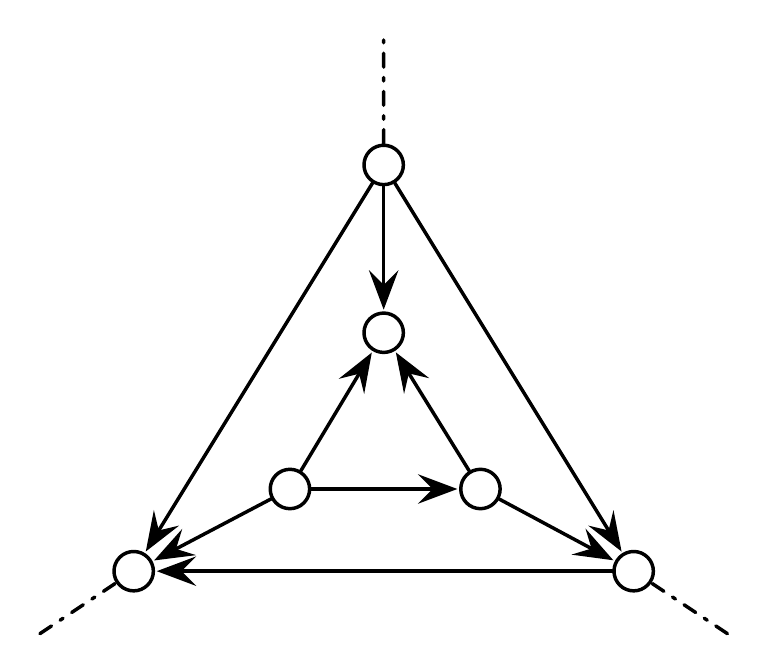}
\caption{\label{fig-gadget-orientation}A good orientation of the element-gadget. It is assumed that the arcs going out of $X_{j}$ occur clockwise from the top of the figure.}
\end{figure}
Because this modification leaves the in-degrees of the vertices in $V - X_{j}$ unchanged, we deduce from Lemma~\ref{lem:jungle} that the new orientation has an entropy strictly smaller than $\vec G$, a contradiction.
\end{proof}

Using Claim~\ref{claim-gadget}, we may assume without loss of generality that $\vec{G}[X_{j}]$ is isomorphic to the orientation of the element-gadget given in Figure~\ref{fig-gadget-orientation}. 
Renaming the vertices if necessary, we may thus suppose that $u_{j,1}$, $u_{j,2}$, $u_{j,3}$ have in-degree $0$, $2$, $3$ in $\vec{G}[X_{j}]$, respectively.

Consider now the three edges $u_{j,1}s_{j_{1}}, u_{j,2}s_{j_{2}}, u_{j,3}s_{j_{3}}$ of $G$. Recall that $j_{1}, j_{2},j_{3}$ denote the indices of the three sets in $\mathcal{S}$ containing the element $u_{j}$. Because $\vec{G}$ has minimum entropy, we may assume that the first of these three edges is oriented out of $X_j$ and the last two toward $X_j$. Indeed, if we have $(u_{j,3},s_{j_{3}}) \in A$ then $\rho_{\vec{G}}(s_{j_{3}})\leq 3$, $\rho_{\vec{G}}(u_{j,3}) = 3$, and changing the orientation of $u_{j,3}s_{j_{3}}$ would decrease the entropy of $\vec{G}$ by Lemma~\ref{lem:jungle}, a contradiction. Moreover, if $(u_{j,2},s_{j_{2}}) \in A$, then re-orienting $u_{j,2}s_{j_{2}}$ either leaves the entropy of $\vec{G}$ unchanged or decreases it. A similar argument holds when $(s_{j_{1}},u_{j,1}) \in A$.

It follows that there is exactly one arc going out of $X_{j}$ in $\vec G$, for every $j \in \{1, \dots, q\}$. We proceed with a second (and last) claim.

\begin{claim}
\label{claim-exact-cover}
Let $m:=|A|$ (\,$= 12q$). Then the entropy of $\vec G$ is at least
$$
\frac{1}{m} \left( 4q \log(m/4) + 7q \log(m/3) +  q \log m  \right),
$$
with equality if and only if there exists an exact cover of $(U, \mathcal{S})$.
\end{claim}
\begin{proof}
As we have seen, we have without loss of generality $\textrm{in-seq}(X_{j})= (4,3,3,1,0,0)$ for every $j \in \{1, \dots, q\}$. Then, each component of $\textrm{in-seq}(S)$, where $S:= \{s_{1}, s_{2}, \dots, s_{q}\}$, is clearly at most 3, and the sum of all of them equals $q$. 
Combining these observations with Lemma~\ref{lem:jungle}, we deduce that the entropy of $\vec G$ is at least the lower bound given in the claim, and that equality occurs if and only if $\textrm{in-seq}(S)$ equals $(3, 3, \dots, 3, 0, 0, \dots, 0)$. We show that the latter happens if and only if there exists an exact cover of $(U, \mathcal{S})$.

Suppose first  $\textrm{in-seq}(S) = (3, 3, \dots, 3, 0, 0, \dots, 0)$, and define $\mathcal{S^{*}} \subseteq \mathcal{S}$ as
$$
\mathcal{S^{*}} := \{ S_{i} : \rho_{\vec G}(s_{i}) > 0\}. 
$$
It is then easily seen that the collection $\mathcal{S^{*}}$ is an exact cover for the set system $(U, \mathcal{S})$.

Now assume that $\mathcal{S^{'}} \subseteq \mathcal{S}$ is an exact cover of $(U, \mathcal{S})$. After permuting the indices $j_1$, $j_2$ and $j_3$, we can assume that the set of $\mathcal{S^{'}}$ containing $u_j$ is $S_{j_1}$. Orienting each edge $s_{i}u_{j,k}$ of $G$ toward $s_{i}$ if $k = 1$, toward $u_{j,k}$ otherwise, and using the orientation of the element-gadgets given in Figure~\ref{fig-gadget-orientation}, we obtain an orientation $\vec G^{*}$ of $G$ where each 
$X_{j}$ has in-degree sequence $(4,3,3,1,0,0)$ and $S$ has in-degree sequence $(3, 3, \dots, 3, 0, 0, \dots, 0)$. Hence, $S$ must also have the same in-degree sequence in $\vec G$, since otherwise $\vec G^{*}$ would have entropy strictly less than $\vec G$, contradicting the optimality of the latter orientation. The claim follows.
\end{proof}

By Claim~\ref{claim-exact-cover}, a polynomial-time algorithm finding
a minimum entropy orientation of $G$ could be used to decide, in polynomial time, if there exists an exact cover of $(U, \mathcal{S})$. This completes the proof of the theorem.
\end{proof}

Let us say that a graph orientation problem has the {\em strict dominance property} if the objective function $F$ to minimize is such that 
$$
F(\vec G) < F(\vec G')
$$
whenever $\vec G$ and $\vec G'$ are two orientations of a graph $G=(V,E)$ such that the in-degree sequence of $V$ in $\vec G$ dominates that of $V$ in  $\vec G'$. Hence, Lemma~\ref{lem:jungle} says exactly that MINEO has the strict dominance property. We remark that, since the proof of Theorem~\ref{th-NPhard} relies solely on that lemma, it follows more generally that every orientation problem with the strict dominance property is NP-hard on planar graphs.

\section{Approximation}
\label{sec-approx}

Throughout this section, we denote by $\OPT(G)$ the minimum entropy
of an orientation of $G$. An orientation of $G$ is {\em biased} if each edge $vw$ with $\deg(v) > \deg(w)$ is oriented toward $v$. It turns out that biased orientations have entropy close to the minimum achievable:

\begin{theorem}
\label{th-approx}
The entropy of any biased orientation of $G$ is at most $\OPT(G) + 1$.
\end{theorem}

Since finding a biased orientation can easily be done in linear time,
Theorem~\ref{th-approx} yields the following corollary:

\begin{cor}
\label{cor-approx}
MINEO can be approximated within an additive error of $1$ bit, in linear time.
\end{cor}

Let $m$ denote the number of edges of $G$ and $d_{v}:= \deg(v) / (2m)$ the normalized degree of a vertex $v$. Given two discrete probability distributions $p$ and $q$ over a common domain $X$, 
we denote by $D(p\parallel q)$ their relative entropy (or Kullback-Leibler distance), 
defined as follows: 
$$
D(p\parallel q) := \sum_{i \in X} p_i \log \frac{p_i}{q_i}.
$$
It is known that $D(p\parallel q)$ is always non-negative (see for instance
Cover and Thomas~\cite[Section 2.6]{TC} for a proof). 

\begin{proof}[Proof of Theorem~\ref{th-approx}.]
Let $\vec G$ be an optimal orientation of $G$. Denote respectively by $p$ and $A$ the in-degree distribution and arc set of $\vec G$. We first rewrite the entropy of $p$ as follows:
\begin{eqnarray*}
\OPT(G) & = & \sum_{v\in V} - p_v \cdot \log p_v\\
& = & \sum_{v\in V} -\frac{\rho_{\vec G}(v)}m \cdot \log \frac{\rho_{\vec G}(v)}m \\
& = & \log m - \frac1m \sum_{v\in V} \rho_{\vec G}(v)\cdot\log\rho_{\vec G}(v) \\
& = & \log m - \frac1m \sum_{(u,v)\in A} \log\rho_{\vec G}(v).
\end{eqnarray*}
Now we observe that for any vertex $v$ we have $\rho_{\vec G}(v) \leq \deg (v)$. We let $\vec G^\flat$ be a biased orientation, $A^\flat$ its arc set, and $p^\flat$ the corresponding in-degree distribution. From our observation, we have
\begin{eqnarray*}
\OPT(G)& \geq & \log m - \frac1m \sum_{(u,v) \in A} \log \max \{ \deg (u), \deg (v)\} \\
& = & \log m - \frac1m \sum_{(u,v) \in A^\flat} \log \deg(v)
\quad \quad \textrm{(because $\vec G^\flat$ is biased)}\\
& = & \log m - \frac1m \sum_{v \in V} \rho_{\vec G^\flat}(v)\cdot \log \deg(v)\\
& = & \sum_{v\in V} -p^\flat_v\cdot \log \frac{\deg (v)}{m} \\
& = & \left( \sum_{v\in V} -p^\flat_v \cdot \log d_v \right) - 1 \\
& = & \left( \sum_{v\in V} -p^\flat_v \cdot \log p^\flat_v\right) + \left( \sum_{v\in V} p^\flat_v \cdot \log \frac{p^\flat_v}{d_v} \right) - 1 \\
& = & H(p^\flat ) + D(p^\flat \parallel d) - 1 \\
\end{eqnarray*} 
Since $D(p^\flat \parallel d)\geq 0$, we have $H(p^\flat ) \leq \OPT (G) + 1$, which concludes the proof.
\end{proof}

We note that the bound given in Theorem~\ref{th-approx} is tight: consider for instance the case where $G$ is a cycle.\\

We end this section with some remarks. For $S \subseteq V$, let $e(S)$ denote the fraction of edges of $G$ incident to a vertex in $S$. Thus $e(V) = 1$. It is well-known that $e$ is a submodular function, that is, satisfies $e(X) + e(Y) \ge e(X \cap Y) + e(X \cup Y)$ for all $X, Y \subseteq V$. We denote the base polytope of $e$ by $P(G)$. 
Letting $p(S) := \sum_{v\in S}p_{v}$ for $S \subseteq V$, 
we thus have
$$
P(G) = \{p \in \mathbb{R}^V : p(S) \le e(S)\ 
\forall S \subseteq V,\ p(V) = 1\}.
$$
It follows from standard results on polymatroids that $P(G)$ is the convex hull of all the in-degree distributions of orientations of $G$;
see for instance Schrijver~\cite{S03-b}. The vertices of $P(G)$ correspond to the acyclic orientations of $G$.

Consider now the following generic linear program, where for each $v\in V$, $c_{v}$ is a fixed non-negative cost:
\begin{equation}
\label{prob-LP-GENERIC}
\begin{array}{rl@{\qquad}l}
\min          &\ds \sum_{v \in V} c_v \cdot p_v\\[2ex]
\textrm{s.t.} &p \in P(G).
\end{array}
\end{equation}
This linear program can be solved by the following greedy algorithm: First order the vertices in $V$ in non-decreasing order of costs, say $v_{1}, v_{2}, \dots, v_{n}$. Then, start with the null vector $p:=(0, 0, \dots, 0)$, and, for each $i=1,\dots,n$, increase the $i$th component of $p$ as much as possible, ensuring that $p(S) \le e(S)$ remains true at all time, for every $S \subseteq V$. It is well-known (see e.g.~\cite{S03-b}) that the resulting point $p$ belongs to $P(G)$, that is, it satisfies also $p(V)=1$, and furthermore that $p$ is an optimal solution to the above linear program.

Let us set $c_{v}:= - \log (\deg(v)/m)$. Thus, we obtain the following linear program:
\begin{equation}
\label{prob-MINEO-LP}
\begin{array}{rl@{\qquad}l}
\min          &\ds \sum_{v \in V} -p_v \cdot \log \frac{\deg(v)}{m} \\[2ex]
\textrm{s.t.} &\ds p \in P(G).
\end{array}
\end{equation}
Since MINEO can be formulated as
\begin{equation}
\label{prob-MINEO}
\begin{array}{rl@{\qquad}l}
\min          &\ds \sum_{v \in V} -p_v\cdot \log p_v\\[2ex]
\textrm{s.t.} &\ds p \in P(G),
\end{array}
\end{equation}
and $-p_v\cdot \log p_v\geq -p_v\cdot \log (\deg(v) / m)$ trivially holds for every $p \in P(G)$,
the optimum value of~\eqref{prob-MINEO-LP} gives a lower bound on $\OPT(G)$.
Now, observe that performing the greedy algorithm to solve~\eqref{prob-MINEO-LP} 
amounts to finding a biased orientation of $G$. Moreover, the in-degree sequence of every such orientation can be produced by the algorithm. 

To conclude, we mention that the natural counterpart of MINEO where one aims at finding an orientation of $G$ with {\em maximum} entropy is polynomial. This is because maximizing a separable concave function over $P(G) \cap \frac1m \mathbb{Z}^{V}$ 
can be done in polynomial time; see~\cite{G91,  H94, MS04}.

\section*{Acknowledgments}

The authors wish to thank Olivier Roussel from the \'Ecole Normale Sup\'erieure de Cachan 
for his preliminary investigations on this topic. We also thank an anonymous referee for an observation that 
made the proof of Theorem~\ref{th-approx} more concise. 

This work was supported by 
the {\em Actions de Recherche Concert\'ees (ARC)\,} fund of the {\em Communaut\'e fran\c{c}aise de Belgique}.

\bibliographystyle{plain}
\bibliography{e-vc}

\end{document}